\documentclass[12pt, a4paper]{article}
\usepackage{graphicx} 

\usepackage{setspace}
\doublespace
\usepackage[utf8]{inputenc}
\usepackage[T1]{fontenc}
\usepackage{graphicx}
\usepackage{xcolor}
\usepackage{enumitem}  

\usepackage[margin=1in]{geometry}
\usepackage{amsmath}
\usepackage{amssymb}

\usepackage{algorithm}%
\usepackage{algorithmicx}%

\usepackage{tabularx}
\usepackage[noend]{algpseudocode}
\usepackage{booktabs}
\usepackage{threeparttable}

\usepackage{url}

\makeatletter
\newcommand{\multiline}[1]{%
  \begin{tabularx}{\dimexpr\linewidth-\ALG@thistlm}[t]{@{}X@{}}
    #1
  \end{tabularx}
}
\makeatother

\usepackage{natbib}
\title{WASP: voting-based ex ante method \\ for selecting joint prediction strategy}

\author{Alicja Wolny-Dominiak, Tomasz Żądło}

\date{}

\begin{document}

\maketitle{}











\begin{abstract}
This paper addresses the topic of choosing a prediction strategy when using parametric or nonparametric regression models. It emphasizes the importance of ex ante prediction accuracy, ensemble approaches, and forecasting not only the values of the dependent variable but also a function of these values, such as total income or median loss. It proposes a method for selecting a strategy for predicting the vector of functions of the dependent variable using various ex ante accuracy measures. The final decision is made through voting, where the candidates are prediction strategies and the voters are diverse prediction models with their respective prediction errors. Because the method is based on a Monte Carlo simulation, it allows for new scenarios, not previously observed, to be considered. The first part of the article provides a detailed theoretical description of the proposed method, while the second part presents its practical use in managing a portfolio of communication insurance. The example uses data from the Polish insurance market. All calculations are performed using the R program.


\vskip 2mm


\end{abstract}

\section{Introduction}

Predictive models are increasingly used in the field of economics to support decision-making and strategic planning. One example is the churn model for companies offering subscription services, such as telecommunications, streaming or Internet services. The model can analyse customer data, such as frequency of service use, satisfaction level, payment history, etc., to predict which customers are more likely to leave. This enables companies to retain customers proactively by offering discounts or personalised offers. The other example is credit scoring in the banking sector. Banks and financial institutions want to assess credit risk before they lend to a customer. Predictive models that analyse financial data, credit history, income, and other factors help assess the risks involved in granting a loan. This allows for more accurate credit decisions. In non-life insurance companies, predictive models can analyse a customer's insurance history, taking into account previous claims, claim counts, payments, as well as other factors that affect risk. Based on this data, the model can predict the value of the risk premium. In macroeconomics, governments and economic institutions want to predict unemployment levels and the country's economic growth rate. Econometric models can analyse a range of macroeconomic data, such as GDP, inflation, and unemployment levels, to predict future trends in the economy. This can be useful for economic policy planning.

In the paper, we consider predictive models in the statistical sense. The relationship between a particular variable (called the dependent variable) and other variables (called independent variables) is assumed. Depending on whether the variable is continuous or multicategorical, one speaks of regression or classification. We focus only on continuous variables, and consequently, we are only considering regression models. The regression predictive models considered can be divided into two main groups. One group contains parametric models, where the primary role is played by the assumed distribution of the random term. The second group consists of nonparametric models, where a specific distribution of the random term is not assumed, making them more flexible but requiring more data for their application. In the area of prediction using regression, various predictive models have been proposed and applied. Of the most popular parametric models, we can mention, for example: the linear model, the generalized linear model (GLM), logistic regression, see \cite{nelder1972glm}. On the other hand, among the non-parametric models, regression trees or support vector machine have been popularised, see \cite{james2013introduction}. Typically, the purpose of using such models is to predict the unknown values of the dependent variable for cross-sectional or longitudinal datasets. In our article, we extend this problem to joint prediction of any functions of the dependent variable. Each function represents a certain characteristic e.g. the quantile or the mean.  A similar problem of the joint estimation of two characteristics -- the at-risk-of-poverty rate and the Gini coefficient based on the parametric Beta regression model assumed for cross-sectional data is studied by \cite{fabrizi_bayesian_2016}. The joint prediction of temperature and relative humidity based on time series (joint prediction of random variables for future periods) using deep learning models is analysed by \cite{gunawan_multivariate_2023}. It should be noted that determining the prediction values by the use of a predictive model requires a specific prediction algorithm. This algorithm is understood as a set of statistical techniques that allows for the prediction of unknown realisations of a random variable based on their known, sample or historical values. What is important, specific predictive algorithms can be applied across various models, while others are tailored to certain models, particularly those with specific optimality properties. Furthermore, the properties of a specific predictive algorithm depend on the model under which they are used. Therefore, we refer it to the term "prediction strategy" as the pair: the predictive model and the prediction algorithm. This distinction between the predictive algorithm and the prediction strategy, commonly understood as the predictive algorithm based on a specific predictive model, is widely employed in the literature, for example by \citet[][p.~14]{van_setten_case-based_2004}, \citet[][p.~286]{vovk_leading_2008}, \citet[][p.~160]{james_introduction_2013}, and \citet[][p.~5]{dang_kalman_2023}. Additionally, this term is similar to the term "estimation strategy", which is defined as the pair: the estimator and the sampling design under which the considered estimator is used \citep[see][p.~26]{cassel_foundations_1977}. For example, BLUP algorithms (Best Linear Unbiased Predictor), proposed by \cite{henderson_estimation_1950} and \cite{royall_linear_1976}, and their estimated versions called the EBLUPs (Empirical Best Linear Unbiased Predictors), can be used only under linear models and only if (\ref{theta}) is a linear combination of the variable of interest. On the other hand, BP algorithms (Best Predictors), studied for example by \cite{molina_small_2010}, and their estimated versions called EBPs (Empirical Best Predictors), can be used to predict any function of the response variable and under a broader class of models, however, under stronger assumptions. Another frequently encountered algorithm is PLUG-IN, see e.g. \cite{boubeta2016empirical}, \cite{hobza_empirical_2016}, \cite{chwila_properties_2022}. This algorithm is definitely more flexible as it can be applied to a variety of models and not just linear or parametric models. In economic applications, it is common to find the GLM or the generalized additive model (GAM) applied to real data. In addition, PLUG-IN also allows the prediction of any characteristic.  It is also important to note that even if the same predictive algorithm is used under two slightly different models belonging to the same narrow class (e.g., two linear models with different combinations of independent variables), it should be considered as two different prediction strategies. 

In applications, the practical problem of selecting the best prediction strategy arises. An imposing solution to this problem is to compare the prediction accuracy measure, e.g. the mean squared error (MSE) or the mean absolute percentage error (MAPE). The approach to determining these measures can be twofold. The first and most commonly used is the ex post approach referring to assessment done after an event has occurred. It involves looking back at what actually happened and comparing observed values of the dependent variable with their corresponding predicted values. Another option is the ex ante approach in which the assessment takes place before an event, e.g. before the realisation of the random variable is known. In this future-oriented approach, at the time the prediction is made, only the predicted values are known. However, the future "true values" are unknown, but it can be generated in a Monte Carlo simulation. The advantage of an ex ante approach is that scenarios which have not occurred before, but could potentially materialise, can be included in this experiment. To simulate the data so-called data generation model can be used playing the role of a real unknown model. This model can be built upon the sample data or it covers specific aspects of the phenomenon being studied that are not reflected in the sample data and cannot be captured by the assumed predictive model. It is worth noting that a future-oriented approach for cross-sectional data corresponds to the prediction of the unobserved variables in an out-of-sample set. In contrast, for longitudinal data it is simply the future as understood on the timeline.

The above considerations lead to an indication of the aim of our article, which is to propose a method for selecting the best strategy for the joint prediction of the assumed characteristics. Furthermore, in the selection, we prefer to consider several accuracy measures simultaneously and not just one and classically select the strategy with the minimum measure. This results in greater prediction precision. Therefore, we propose to use voting as it is in classification, \citep[e.g.][]{bauer_empirical_1999, burka_voting_2022, daneshvar_voting-based_2023}. We show that in the case of prediction based on regression models, the proposed criteria for selection can be based on various voting algorithms inspired by the social choice literature, see e.g. \cite{balinski_theory_2007}, \cite{grover2019polarization}.  

Summing up, our paper aims to propose a voting-based ex ante method for selecting a joint prediction strategy, which is characterized by the following important features:
\begin{itemize}
    \item joint prediction of any vector of characteristics,
    \item using a set of any ex ante prediction accuracy measures, in which future unknown values are obtained in a Monte Carlo experiment using a data generation model,
    \item utilizing any type of data (cross-sectional, longitudinal, time-series),
    \item considering various future scenarios defined by any class of models (both parametric and nonparametric models can be used simultaneously).
\end{itemize}
In the further part of this article, we refer to the method briefly as the WASP method - Voting-based ex Ante for Selecting joint Prediction strategy method. The remainder of the paper is structured as follows: Section 2 explores in details the problem of model-based prediction, the definition of the prediction strategy and fundamental differences between the ex ante and the ex post approaches. Section 3 introduces the subsequent steps of the WASP method. In Section 4, a practical example is presented to demonstrate the operation of the WASP method in the prediction of total claims and median of claims in the automobile insurance portfolio. Section 5 includes the conclusions.

\section{Theoretical foundations of prediction in an ex ante approach}
To understand the nature and operation of the proposed WASP selection method, it is necessary to take a broad view of the different areas of statistical modelling. In this section, we present the theoretical background of the main areas applicable to the WASP method. We start with basic notation and introduce the problem of predicting an arbitrary vector function of the response variable for in-sample and out-of-sample data. We then present the general form of the predictive model. Finally, we discuss the ex ante approach to accuracy assessment and the data generation algorithm. 

\subsection{Input data} \label{subsec: input data}
In our approach, we assume that the following information is available for the set $S$ of $n$ sample observations (see the top-left part of Figure \ref{fig_approaches}): the matrix of $q$ independent variables denoted by $\mathbf{X}_S = [\begin{matrix} x_{ij} \end{matrix}]_{n \times q}$ and the vector of values of the response (dependent) variable $\mathbf{y}_S = [\begin{matrix} y_1 & y_2 & \dots & y_n \end{matrix} ] ^T$. What is important, $\mathbf{X}_S$ is fixed (non-random) while $\mathbf{y}_S$ is treated as a realisation of the random vector $\mathbf{Y}_S = [\begin{matrix} Y_1 & Y_2 & \dots & Y_n \end{matrix} ] ^T$. Let the size of the out-of-sample observations set,  denoted by $R$  (see the top-right part of Figure \ref{fig_approaches}), be denoted by $k$. The known or assumed, fixed (non-random) matrix of independent variables for this set is denoted by $\mathbf{X}_R = [\begin{matrix} x_{ij} \end{matrix}]_{k \times q}$. The random vector of the dependent variable for out-of-sample observations, whose realisations are unknown, is denoted by $\mathbf{Y}_R = [\begin{matrix} Y_1 & Y_2 & \dots & Y_k \end{matrix} ] ^T$. Let $\mathbf{Y} = [\begin{matrix} \mathbf{Y}_S^T & \mathbf{Y}_R^T \end{matrix} ] ^T$ and $\mathbf{X} = [\begin{matrix} \mathbf{X}_S^T & \mathbf{X}_R^T \end{matrix} ] ^T$. The introduced notation is a standard setup both in classic econometrics \citep[e.g.][chapter 2]{greene_econometric_2012} and in machine learning \citep[e.g.][p.~16]{james_introduction_2013}. 

The following considerations apply to several types of data. First, it can be used for cross-sectional data, which involves observing $n$ sample elements while the entire population of size $(n+k)$ is of interest. Second, it can be employed in time series analysis where, based on $n$ observations of the dependent variable, values for $k$ future periods are predicted. Finally, this notation also covers longitudinal data analysis. However, the out-of-sample set includes in this case out-of-sample observations for current and past periods, as well as all population observations in future periods.

\subsection{Predictive Model}
Let us define the random vector $\mathbf{Y} = [\begin{matrix} \mathbf{Y}_S^T & \mathbf{Y}_R^T \end{matrix} ] ^T$ of the response (dependent) variable for both sets -- sample and out-of-sample observations \citep[compare][]{valliant_finite_2000}. We consider the problem of  joint prediction of $C$  given functions of the vector $\mathbf{Y}$ of the dependent variable, 
\begin{equation} \label{theta.vector}
{\pmb\theta}  = \left[ \begin{matrix} {\theta^{(1)}} (\mathbf{Y})  & {\theta^{(2)}}(\mathbf{Y}) & \dots &{\theta^{(C)}}(\mathbf{Y})   \end{matrix} \right]^T,
\end{equation} where 
\begin{equation} \label{theta}
{\theta^{(c)}} (\mathbf{Y}) = {\theta^{(c)}} \left( \left[ \begin{matrix} \mathbf{Y}_S \\ \mathbf{Y}_R \end{matrix} \right]\right), c=1,...,C.
\end{equation}
In the general case, the function $\theta^{(c)}$  may correspond to any population and subpopulation characteristics such as quantiles and variability measures. It could also be a linear combination of $\mathbf{Y}$, e.g. a mean or one future realisation of the response variable allowing to model time series. The prediction of (\ref{theta.vector}) rests on certain assumptions made about the distribution of $\mathbf{Y}$, which is referred to as the model. In further consideration, various regression models, denoted by $M(\mathbf{Y},\mathbf{X})$, is considered. An example of a regression model can be written as follows:
\begin{equation}
	\label{model}
	\left\{ \begin{array}{c}
		\mathbf{Y}=h(\mathbf{X}) + \pmb{\xi} \\
		E(\pmb{\xi})= \mathbf{0} \\
            Var(\pmb{\xi})= \mathbf{V}
  	\end{array} \right.
\end{equation}
where $h(.)$ is a fixed but unknown function of independent variables, $\pmb{\xi}$ is a random term with $\mathbf{0}$ mean and unknown variance-covariance matrix $\mathbf{V}$. The formula (\ref{model}) covers many models considered both in classic econometrics and in machine learning. Special cases of (\ref{model}), for example, include the multiple regression model \cite[p.~87]{baltagi_econometrics_2021} and the linear mixed model \cite[p.~2]{jiang_linear_2007}. What is more, a special case of model (\ref{model}), where the independence of elements of $\pmb{\xi}$ is additionally assumed, is also considered in machine learning \citep[p.~28]{hastie_elements_2009}. 

The unknown function of independent variables in (\ref{model}) is denoted by $h(.)$. For the precise presentation of our proposal, let us introduce additional notations. Let $\hat{H}^{(l)}(.)$, as a function of $\mathbf{Y}_S$ and  $\mathbf{X}_S$, be an estimator of $h(.)$ based on $l$th predictive model $M^{(l)}(\mathbf{Y},\mathbf{X})$. Moreover, $\hat{h}^{(l)}(.)$, obtained based on  $\mathbf{y}_S$ (a realisation of $\mathbf{Y}_S$) and  $\mathbf{X}_S$, be the estimate --  the realisation of $\hat{H}^{(l)}(.)$. It is worth noting that this notation is consistent with the classic notation used in mathematical statistics, where, for example, the unknown population mean is denoted by $\mu$, its estimator by $\bar{X}$ and its value by $\bar{x}$. Finally, if the $l$th predictive model is either parametric or nonparametric, the estimates of $h(.)$ is denoted by $\hat{h}^{(l)}_{PAR}(.)$ or $\hat{h}^{(l)}_{NPAR}(.)$, respectively.

We make a distinction between the parametric and nonparametric models, as discussed in \citet[][p.~21--24]{james_introduction_2013}. In parametric models, denoted by $M_{PAR}(\mathbf{Y},\mathbf{X})$, a parametric form of $h(\mathbf{X})$ is assumed. For example, in linear models $h(\mathbf{X}) = \mathbf{X} \boldsymbol{\beta}$, where $\boldsymbol{\beta}$ is a vector of unknown parameters. Then, these parameters are estimated instead of the function itself. This approach is usually simpler than estimating the undetermined function $h(\mathbf{X})$. However, the chosen parametric form of the function may not accurately approximate the unknown, complex form of $h(\mathbf{X)}$. Nonparametric models, denoted by $M_{NPAR}(\mathbf{Y},\mathbf{X})$, do not require an explicit specification of the functional form of $h(\mathbf{X})$. This flexibility allows nonparametric approaches to capture a wider range of functional relationships between variables, which parametric models may miss. Nonetheless, nonparametric methods are associated with a significant limitation: since they do not reduce the estimation of $h(\mathbf{X})$ to a small number of parameters, a substantially larger sample size is necessary to obtain a reliable estimate of $h(\mathbf{X})$ compared to the sample size required by parametric methods. Therefore, we consider both approaches in our further analyses, taking into account their respective advantages and disadvantages. It is noteworthy that in our proposal, several models (both parametric and nonparametric) selected by the user can be incorporated into the procedure, thereby enabling the consideration of diverse distributions and scenarios of the changes of the response variable in space or time.

\subsection{Prediction algorithm}
After defining the various characteristics of interest given by the general vector formula (\ref{theta.vector}) and specifying various models that can describe the distribution of $\mathbf{Y}$, the next step is to determine the predictive algorithms. For simplicity of presentation, let us limit the further considerations to the PLUG-IN predictive algorithm mentioned in the introduction. It is a very flexible algorithm because it can be used to predict any characteristic using any parametric or nonparametric model. Moreover, as shown in studies presented, for example, by \citet[][p.~20]{chwila_properties_2022}, it can be more accurate (or have similar accuracy) compared with the EBP algorithms (Empirical Best Predictor). In the case of predicting one characteristic (\ref{theta}) under the $l$th predictive model the PLUG-IN can be written as follows:
\begin{equation} \label{etheta}
\hat{\theta}_{PLUG-IN}^{(l)} = \hat{\theta}_{PLUG-IN}^{(l)}(\mathbf{Y}_S)  = {\theta} \left( \left[ \begin{matrix} \mathbf{Y}_S \\ \hat{H}^{(l)}(\mathbf{X}_R) \end{matrix} \right]\right).
\end{equation}
In the case of predicting a vector of $C$ characteristics (\ref{theta.vector}) under the $l$th predictive model it is given by:
\begin{equation} \label{etheta.vector}
\hat{\pmb\theta}^{(l)}_{PLUG-IN}  = \left[ \begin{matrix} 
\hat{\theta}^{(l,1)}_{PLUG-IN}(\mathbf{Y}_S) &
\hat{\theta}^{(l,2)}_{PLUG-IN}(\mathbf{Y}_S) &
\dots
\hat{\theta}^{(l, C)}_{PLUG-IN}(\mathbf{Y}_S) 
\end{matrix} \right]^T,
\end{equation}
where $\hat{H}(.)$ is an estimator of $h(.)$ (see (\ref{model})). Finally, when considering the PLUG-IN predictive algorithm (\ref{etheta.vector}) and $L$ predictive models (i.e., $l=1,2,\dots, L$), it leads to $L$ potential prediction strategies: 
\begin{equation} \label{Lstrategies}
\hat{\pmb\theta}^{(1)}_{PLUG-IN}, \hat{\pmb\theta}^{(2)}_{PLUG-IN}, \dots,\hat{\pmb\theta}^{(l)}_{PLUG-IN}, \dots, \hat{\pmb\theta}^{(L)}_{PLUG-IN}.
\end{equation}

\subsection{Two approaches to prediction accuracy}
\begin{figure} 
	\begin{center}
		\includegraphics[width=15cm]{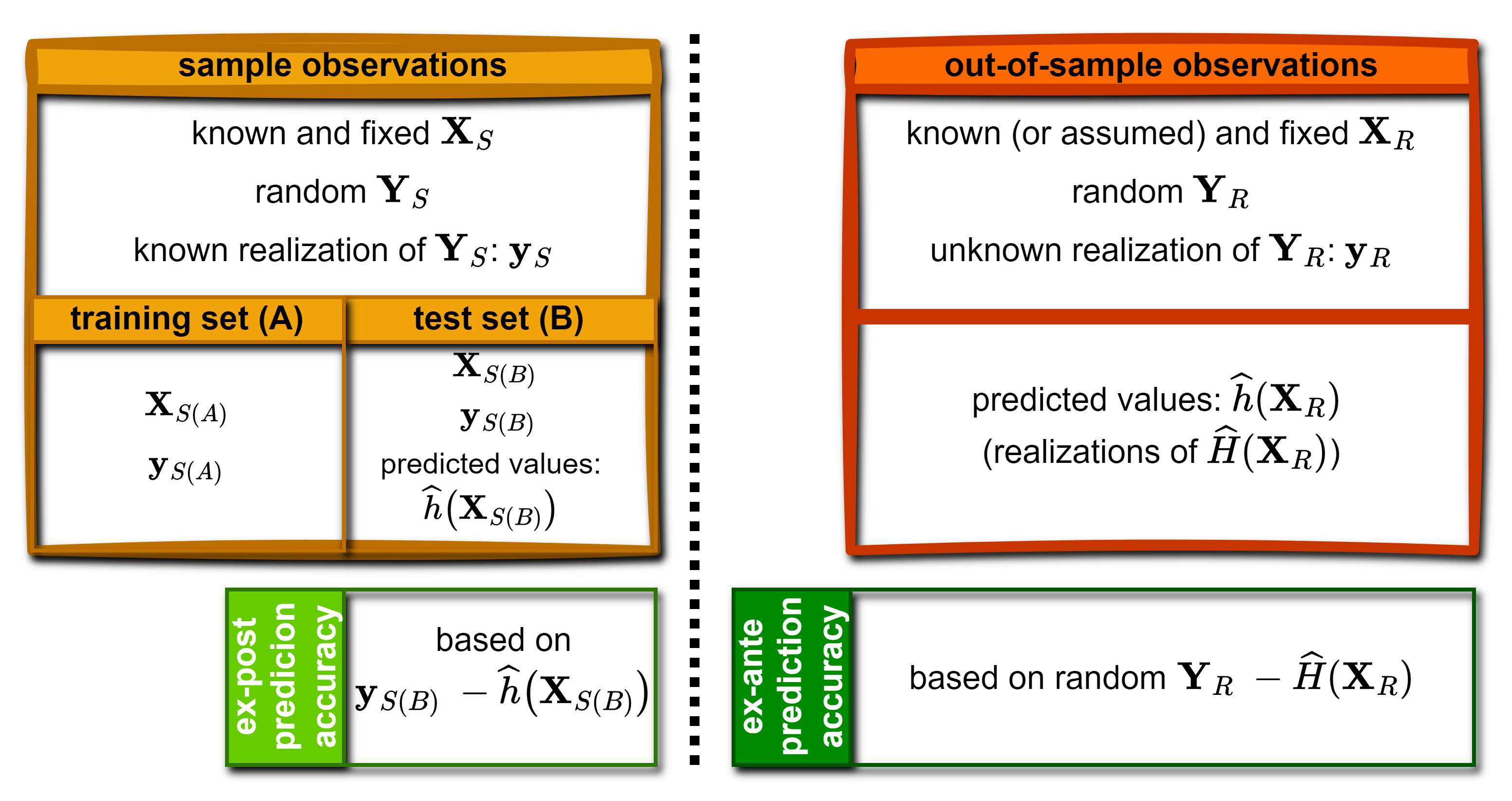} 
		\caption{Approaches to prediction assessment}
  	\label{fig_approaches}
	\end{center}
\end{figure}

The assessment of the prediction accuracy is usually considered within two procedures presented in Figure \ref{fig_approaches}. They involve the assessment of the ex-post prediction accuracy and, as considered in our proposal, the ex ante prediction accuracy.

Firstly, in machine learning, the ex-post prediction accuracy is usually conducted, called shortly the prediction accuracy as in \citet[p.~52]{hastie_elements_2009}. In this case, based on a model fitted on the training set and auxiliary variables available for the test set, predicted values $\hat{h}(\mathbf{X}_{s(B)})$, where $i=1,2,\dots, n_B$, are computed for the test set (denoted by $B$ in Figure \ref{fig_approaches}). Because real values of the variable of interest ($y_i$ for $i=1,2,\dots, n_B$) are available for the test set, the comparison of $\hat{h}(\mathbf{X}_{s(B)})$ and $y_i$, for $i=1,2,\dots, n_B$, is possible. The mean (over the test set) of squared differences between these values is called in machine learning the test MSE \cite[see][p.~30]{james_introduction_2013}. What is more, as discussed in \citet[pp.~241-249]{hastie_elements_2009}, the preferable procedure, in this case, is the K-Fold Cross-Validation, where the division of the sample into the training and the test sets is performed K-times averaging ex-post accuracy measures over the executed K-steps. It is important to note that the ex-post accuracy assessment is not considered for random variables but is based on comparing their realisations -- the real values of the response variable, with their predicted values. This implies that conclusions based on this approach strongly depend on the assumption that the random process considered will be the same for out-of-sample data. Thus, the ex-post prediction accuracy assessment does not allow to take into account processes or events not observed in the sample dataset.

Secondly, the ex ante prediction accuracy can be assessed. In the out-of-sample set (denoted by $R$ in Figure \ref{fig_approaches}), the realisations of random variables $Y_i$, for $i=1,2,\dots, k$, which form a vector $\mathbf{Y}_R$, are not known. In this approach, in contrast to the ex-post accuracy assessment, instead of comparing the real realisations of random variables, characteristics of the distribution of the difference $\mathbf{Y}_R - \hat{H}(\mathbf{X}_R)$ are of interest. In simple cases, the prediction mean squared error MSE can be derived using analytical methods. Then, by replacing the unknown model parameters with estimators, we can obtain a prediction MSE estimator. In more complex cases, bootstrap methods are used as a convenient but time-consuming solution, with proven properties in some cases \citep[e.g.][]{butar_measures_2003, zadlo_accuracy_2020}. However, to assess the accuracy, we rely on one assumed model. In standard applications, ex ante prediction accuracy measures are estimated for a specific prediction algorithm under the assumed model. Hence, in the case of using various prediction strategies in one analysis, as a result of the analysis we obtain values of estimated ex ante prediction accuracy measures under different models, which are not directly comparable. 

Our proposed method is a generalisation of the standard ex ante prediction assessment because various models and scenarios of the generation of out-of-sample observations is taken into account for all prediction strategies. Hence, it solves the problem of an unknown real-life realisation of the vector of the response variable for out-of-sample observations (denoted in Figure \ref{fig_approaches} by $\mathbf{y}_R$) by generating several times realisations of a random vector $\mathbf{Y}_R$ under different assumptions. Moreover, these scenarios could include unlikely changes in the response variable distribution, which could, however, significantly impact the decisions made by data users such as policy-makers or entrepreneurs.

\subsection{Ex ante prediction accuracy measures}
In the above description and Figure \ref{fig_approaches}, we have focused on a common problem of comparing random variables (or their real realisations) with random results obtained via a prediction algorithm (or obtained predicted values), one by one on a unit level. Afterwards, we calculate a measure summarizing all these unit-level comparisons, such as the mean of squared differences. However, as shown in equation (\ref{theta}), the problem we consider is not necessarily the prediction of a single random variable for a future period or one unobserved population element. Instead, it involves predicting any function of $\mathbf{Y}$, which includes various population and subpopulation characteristics. To assess the accuracy of a prediction strategy, represented by $\hat \theta = \hat \theta(\mathbf{Y}_S)$, of a characteristic $\theta = \theta(\mathbf{Y})$, firstly, the prediction error is defined as $U=\hat{\theta}-\theta$. Therefore, the well-known prediction RMSE, measuring ex ante prediction accuracy, has the following formula in the considered case:
\begin{equation}\label{RMSE}
	RMSE(\hat{\theta})=\sqrt{E(\hat{\theta}-\theta)^{2}}=\sqrt{E({{U}^{2}})}.
\end{equation}
The alternative to the RMSE based on the mean could be the QAPE based on quantiles. It represents the $p$th quantile of the absolute prediction error $|U|$, see \cite{zadlo_parametric_2013} and  \cite{wolny-dominiak_bootstrap_2022}, and it is given by:
\begin{equation}\label{QAPE}
	QAPE_p(\hat{\theta}) = \inf \left\{ {x:P\left( {\left| {{\hat{\theta}-\theta}} \right| \le x} \right) \ge p} \right\} =\inf \left\{ {x:P\left( {\left| {{U}} \right| \le x} \right) \ge p} \right\}
\end{equation}
This measure informs that at least $p100$\% of observed absolute prediction errors are smaller than or equal to $QAPE_p(\hat{\theta})$, while at least $(1-p)100\%$ of them are higher than or equal to $QAPE_p(\hat{\theta})$. These quantiles provide information about the relationship between the magnitude of the error and the probability of its occurrence. Using the QAPE makes it possible to obtain a comprehensive description of the distribution of prediction errors, which is impossible by relying solely on the average (measured by the RMSE). Additionally, the MSE represents the average of squared prediction errors, which are usually positively skewed. Therefore, the mean should not be used as a measure of central tendency for such distributions. The described accuracy prediction measures RMSE and QAPE can be obtained through the Monte Carlo simulation experiment under different models, allowing the comparisons of prediction strategies under different scenarios. They can be computed based on equations (\ref{RMSE}) and (\ref{QAPE}), where prediction errors are replaced by their values generated in the Monte Carlo simulation experiment.

\subsection{Data generation process} \label{sec:boot}
The WASP method applies the Monte Carlo simulation experiment under different data generation models. In fact, it corresponds to the parametric bootstrap algorithm to generate values of the response variable under parametric or nonparametric models. In this algorithm, data generation models are denoted by $M_{(g)}$, where $g=1,2,\dots, G$ . These models should not be confused with predictive models, even though they may be of the same class e.g. GLMs. The data generation process allows to compute the prediction accuracy measures (\ref{RMSE}) and (\ref{QAPE}) under any model chosen by the user. It is possible by replacing prediction errors $U=\hat{\theta}-\theta = \hat \theta(\mathbf{Y}_S) - \theta(\mathbf{Y})$ in (\ref{RMSE}) or (\ref{QAPE}) by their generated bootstrap realisations $U_{gen}= \hat \theta(\mathbf{y}_{s \,  gen}) - \theta(\mathbf{y}_{gen})$, where $\mathbf{y}_{s \,  gen}$ and  $\mathbf{y}_{gen}$, are generated sample and population vectors of the dependent variable, respectively. 

If the model is parametric, the parametric bootstrap procedure, studied, for example, by \citet{gonzales2007} and \citet{gonzales2008}, can be used to generate $\mathbf{Y}$ vector. It means that a parametric distribution is assumed for $\mathbf{Y}$, the parameters are estimated based on sample data, namely $\mathbf{y}_s$ and $\mathbf{X}_s$, and bootstrap realisations of $\mathbf{Y}$ are generated from the estimated distribution using both $\mathbf{X}_s$ and $\mathbf{X}_R$. If regression model (\ref{model}) is parametric, it implies that we estimate parameters, $h(.)$ is the function of. Then, we obtain estimate $\hat{h}_{PAR}(.)$ and the fitted values $\hat{h}_{PAR}(\mathbf{X})$ computed for both the sample and out-of-sample elements. Additionally, we also estimate the unknown parameters of the distribution of the random term $\pmb{\xi}$ under (\ref{model}). The generated under the $g$th model (where $g=1,2,\dots, G$) the $b$th (where $b=1,2,\dots, B$) parametric bootstrap realisation of $\mathbf{y}$ vector  is given by:

\begin{equation} \label{parGen}
\mathbf{y}_{gen}^{(g,b)} = \hat{h}^{(g)}_{PAR}(\mathbf{X}) + \pmb{\xi}_{gen}^{(g,b)},
\end{equation}
where $\hat{h}^{(g)}_{PAR}(.)$ is a parametric estimate of $h(.)$ obtained based on the original sample under the assumption of the $g$th (here: parametric) data generation model, $\pmb{\xi}_{gen}^{(g,b)}$ is a generated $b$th realisation of an error term from the estimated parametric distribution assumed for $\pmb{\xi}$ under the $g$th model. For example, under (\ref{model}) and the additional normality assumption, random terms are realisations of the multivariate normal distribution $N(\mathbf{0},\hat{\mathbf{V}})$, where $\hat{\mathbf{V}}$ is an estimate of the variance-covariance matrix of random $\pmb{\xi}$ denoted in (\ref{model}) by $\mathbf{V}$. 

However, the assumed model does not have to be a parametric one. Hence, let us propose a method of the generation of $\mathbf{Y}$ vector based on a nonparametric model. The method is inspired by the residual bootstrap procedure used for parametric models \citep{carpenter_novel_2003, chambers_random_2013, thai_comparison_2013}. However, instead of a parametric model, we use a nonparametric model, and instead of the simple random sampling of residuals, we propose a generation of the random term based on a nonparametric estimate of the distribution fitted to residuals. We consider kernel density estimation, studied among others by \citet{fauzi_statistical_2023}. Let $\hat{h}_{NPAR}(.)$ be a nonparametric estimate of $h(.)$ in (\ref{model}). Then, the vector of residuals based on sample observations can be defined as $\mathbf{r}_S = \mathbf{y}_S - \hat{h}_{NPAR}(\mathbf{X}_S)$. Finally, the proposed bootstrap procedure for nonparametric models is as follows. Under the $g$th model (where $g=1,2,\dots, G$) generated $b$th ($b=1,2,\dots, B$) nonparametric bootstrap realisation of $\mathbf{y}$ vector is given by:
\begin{equation} \label{nonparGen}
\mathbf{y}_{gen}^{(g,b)} = \hat{h}^{(g)}_{NPAR}(\mathbf{X}) +  \bar{gen}^{(b)}(\hat{f}(\mathbf{r}^{(g)}_S), (n+k)),  
\end{equation}
where $\hat{h}^{(g)}_{NPAR}(.)$ is a nonparametric estimate of $h(.)$ based on the original sample under the assumption of the $g$th (here: nonparametric) data generation model, $\hat{f}(\mathbf{r}^{(g)}_S)$ is a kernel density estimate of the distribution of residuals $\mathbf{r}^{(g)}_S$ computed under the $g$th model, and $\bar{gen}^{(b)}(\hat{f}(\mathbf{r}^{(g)}_S), (n+k))$ is a $(n+k) \times 1$ zero centred vector of values generated based on kernel estimate $\hat{f}(\mathbf{r}^{(g)}_S)$ in the $b$th iteration. 

Note that in the method WASP method we assume that the realizations of $\mathbf{Y}$ are based on data generation models that mimic sample data, i.e., estimated based on sample data. However, the data generation process may apply any model chosen by the user, including models that cover features not observed in sample data, for example expert models \citep{franses_expert_2009}.

\section{The proposed WASP method}
\begin{figure}
	\begin{center}
		\includegraphics[width=15cm]{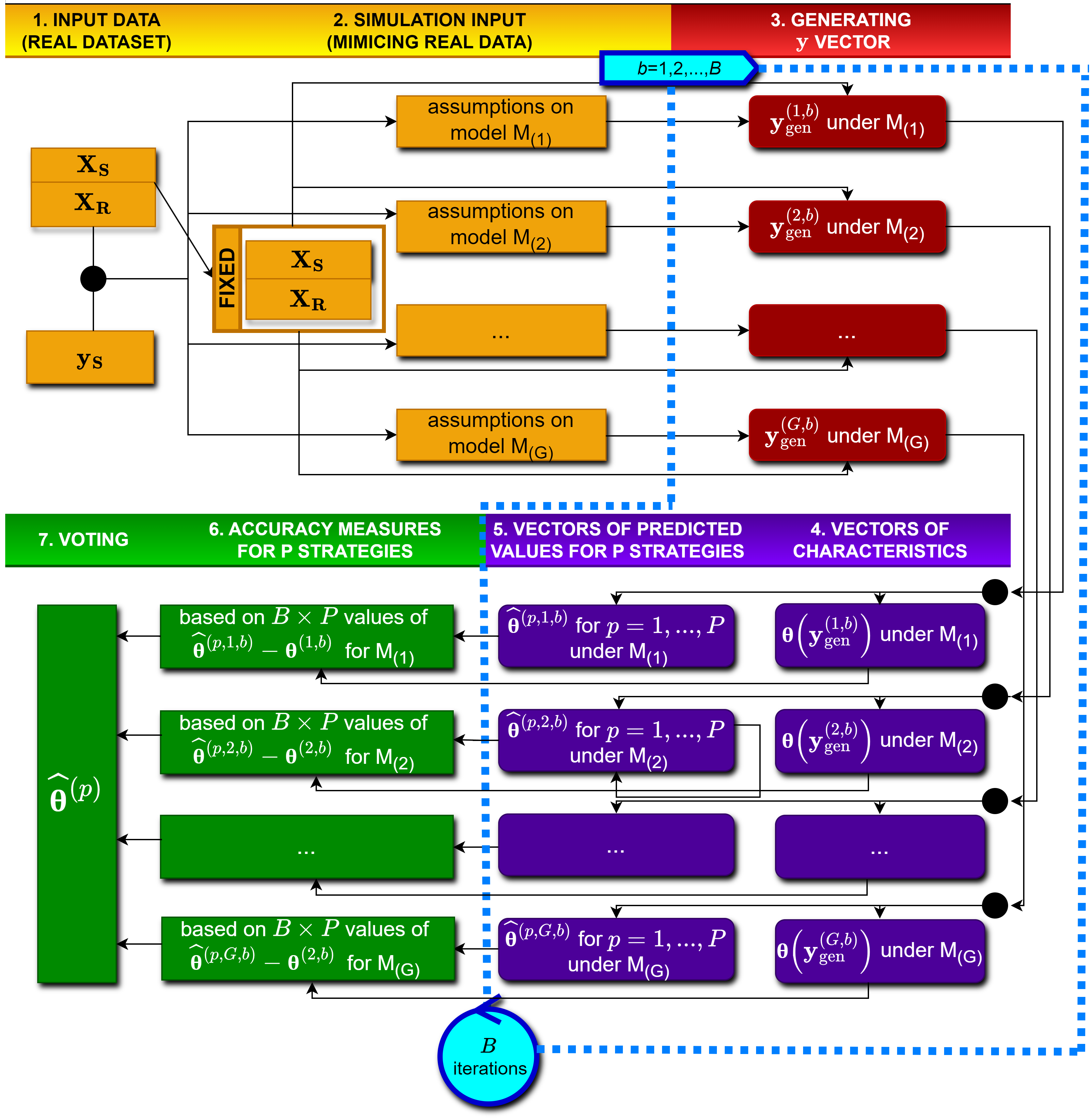} 
		\caption{ Proposed WASP method}
  	\label{fig_scheme}
	\end{center}
\end{figure}

Our aim is to choose the best prediction strategy in terms of the ax ante prediction accuracy measured over various predictive models. It is conducted in the following four steps discussed in subsequent subsections:
\begin{enumerate}
    \item preparing the Monte Carlo experiment input (see points 1 and 2 in Figure \ref{fig_scheme}),
    \item  generation of the dependent variable values (see point 3 in Figure \ref{fig_scheme}),
    \item prediction (see points 4 and 5 in Figure \ref{fig_scheme}),
    \item the ex ante prediction accuracy assessment and the selection of the prediction strategy based on voting algorithms (see points 6 and 7 in Figure \ref{fig_scheme}).
\end{enumerate}

\vskip 3mm

The WASP method is based on a certain similarity between the problem under analysis and the problem of voters selecting one candidate out of many. In our proposal, potential prediction strategies play the role of candidates (compare columns of Table \ref{tab_accuracy measures matrix} and Table~\ref{tab_voting matrix}), and simulation scenarios play the role of voters (compare rows of Table \ref{tab_accuracy measures matrix} and Table \ref{tab_voting matrix}). 
In the considered ex ante approach, we involve various simulation scenarios to choose a prediction strategy, which can be seen as various voters assessing candidates to select one of them. In contrast, in the ex-post approach, the prediction strategies are compared once with real values. It can be interpreted as one voter selecting a candidate. 

It is essential to note that while voting results are commonly measured on an ordinal scale (see Subsection \ref{subsec. voting}), our results, defined as values of ex ante prediction accuracy measures, are expressed on a ratio scale. However, these two approaches are not very different. Voting results can be considered, as in \cite{green-armytage_statistical_2016}, as outcomes based on voters' real-valued utility functions, which are not observed directly. Our results are also real-valued, but, in contrast, they are available in practice. Due to this similarity, we refer it to the matrix of transformed values of accuracy measures (not necessarily to values on an ordinal scale), which are utilized to select the prediction strategy, as the voting matrix.

\begin{table}
\caption{Example of a matrix of values of accuracy measures for $P$ prediction strategies under $S$ simulation scenarios}
\begin{center}
\begin{tabular}{c | c c c c c c | } \label{tab_accuracy measures matrix}
 & strategy 1 & strategy 2 & \dots & strategy $P$ \\
\hline
{simulation scenario 1} & $r_{11}$ &  $r_{12}$ &  \dots & $r_{1P}$ \\
{simulation scenario 2} & $r_{21}$ &  $r_{22}$ &  \dots & $r_{2P}$ \\
\dots  & \dots & \dots &  \dots  \\
{simulation scenario S} & $r_{S1}$ &  $r_{S2}$ &  \dots & $r_{SP}$ \\
\hline
\end{tabular}
\end{center}
\end{table}

\begin{table}
\caption{Example of a voting matrix for $S$ voters choosing one of $P$ candidates}
\begin{center}
\begin{tabular}{c | c c c c c c | } \label{tab_voting matrix}
 & candidate 1 & candidate 2 & \dots & candidate $P$ \\
\hline
{voter 1} & $v_{11}$ &  $v_{12}$ &  \dots & $v_{1P}$ \\
{voter 2} & $v_{21}$ &  $v_{22}$ &  \dots & $v_{2P}$ \\
\dots  & \dots & \dots &  \dots  \\
{voter S} & $v_{S1}$ &  $v_{S2}$ &  \dots & $v_{SP}$ \\
\hline
\end{tabular}
\end{center}
\end{table}

\subsection{Monte Carlo experiment input}
Let us consider the first step of the WASP method, which consists of points 1 and 2 presented in Figure \ref{fig_scheme}. Using the notation presented in Subsection \ref{subsec: input data}, it is assumed that our input data consists of: sample matrices of values of the dependent variable $\mathbf{y}_S$ and independent variables $\mathbf{X}_S$ and the out-of-sample matrix of values of independent variables $\mathbf{X}_R$ (see point 1 in Figure \ref{fig_approaches}).
The values of the independent variables are treated as fixed (non-random), while the real values of the dependent variable are considered as realisations of random variables. In the next step (see point 2 in Figure \ref{fig_approaches}), different assumptions on the distribution of the random vector $\mathbf{Y}$ are considered in the form of parametric and nonparametric models. We estimate the parameters of the parametric models and fit the nonparametric models based on the sample data. For (\ref{model}), this means that $h(.)$ is estimated based on either a parametric or nonparametric framework. This stage also allows us to define any model that explains changes in the response variable, even ones that cover phenomena not observed in the sample. This allows decision-makers to consider non-observed factors or events that may be of interest to them.

\subsection{Data generation}
The second step of the procedure, which consists of point 3 presented in Figure \ref{fig_scheme}. In the following considerations, we assume that based on the known sample realisations of the dependent variable $\mathbf{y}_S$ and sample values of independent variables $\mathbf{X}_S$ the data generation model is fitted, and then, additionally using out-of-sample values of independent variables $\mathbf{X}_R$, model realisations of $\mathbf{Y} = [\begin{matrix} \mathbf{Y}_S^T & \mathbf{Y}_R^T \end{matrix} ] ^T$ are generated. This step is similar to the model-based bootstrap data generation as considered in \citet[][subchapters 2.4.2 and 6.3]{chernick_introduction_2014}. However, values of the response variable are not generated only for the sample but also for out-of-sample elements. Moreover, in relatively rare (from the practical point of view) cases not considered in this paper, the realisation of $\mathbf{Y}$, denoted by $\mathbf{y} = [\begin{matrix} \mathbf{y}_S^T & \mathbf{y}_R^T \end{matrix} ] ^T$ is known. It is possible, for example, for census data or secondary datasets presented for a specific aggregation level and published by statistical offices. In such a case, the considered data generation model can be fitted based on both sample and out-of-sample data, namely $\mathbf{y} = [\begin{matrix} \mathbf{y}_S^T & \mathbf{y}_R^T \end{matrix} ] ^T$ and $\mathbf{X} = [\begin{matrix} \mathbf{X}_S^T & \mathbf{X}_R^T \end{matrix} ] ^T$. Hence, the data generation process under a model fitted for such a dataset is not a bootstrap procedure but a Monte Carlo simulation study designed for the whole set, not only the sample. To generate realisations of $\mathbf{Y}$ under a parametric model, we use a parametric bootstrap and formula (\ref{parGen}). The generation process under nonparametric models is based on our proposal given by (\ref{nonparGen}). Moreover, these population data generation proposals mimicking the sample can be expanded to incorporate user's conceptions of the studied phenomenon not observed in the data.

\subsection{Prediction}

Considering the third step of the procedure, we focus on points 4 and 5 presented in Figure \ref{fig_scheme}.  Assuming without loss of generality, that the first $n$ elements of $\mathbf{y}_{gen}^{(g,b)}$ are for the sample elements, we can decompose it as follows: $\mathbf{y}_{gen}^{(g,b)} = \left[ \begin{matrix} \mathbf{y}_{s \, gen}^{(g,b)} \\ \mathbf{y}_{r \, gen}^{(g,b)} \end{matrix} \right]$, where $\mathbf{y}_{s \,  gen}^{(g,b)}$ is the vector of values of the response variable generated for the sample and  $\mathbf{y}_{r \,  gen}^{(g,b)}$ for out-of-sample elements. 

We consider the problem of computation of the characteristics of interest and their predicted values based on simulation results.  Based on generated vectors of the response variable $\mathbf{y}_{gen}^{(g,b)}$ for each out of $G$ data generation models and each out of $B$ iterations for both the sample and out-of-sample observations, we compute vectors of $C$ characteristics given by (\ref{theta.vector}) as $\pmb\theta(\mathbf{y}_{gen}^{(g,b)}) = \left[ \begin{matrix} 
{\theta}_{(1)}(\mathbf{y}_{gen}^{(g,b)}) &
{\theta}_{(2)}(\mathbf{y}_{gen}^{(g,b)}) &
\dots
{\theta}_{(C)}(\mathbf{y}_{gen}^{(g,b)}) 
\end{matrix} \right]^T,$
where $g=1,2,\dots, G$ and $b=1,2,\dots, B$. Then, each vector of characteristics is predicted using $P$ considered prediction strategies where in general case $L$ predictive models can be used, see (\ref{Lstrategies}). In special case of one prediction algorithm and $L$ predictive models, $P = L$. It means that in each case, based on the sample subvector of $\mathbf{y}_{gen}^{(g,b)}$, denoted by $\mathbf{y}_{s \, gen}^{(g,b)}$, we predict $\pmb\theta(\mathbf{y}_{gen}^{(g,b)})$ using 
$\hat{\pmb\theta}^{(p,g,b)} = \hat{\pmb\theta}^{(p)}(\mathbf{y}_{s \, gen}^{(g,b)}) = \left[ \begin{matrix} 
\hat{\theta}^{(p)}_{(1)}(\mathbf{y}_{s \, gen}^{(g,b)}) &
\hat{\theta}^{(p)}_{(2)}(\mathbf{y}_{s \, gen}^{(g,b)}) &
\dots &
\hat{\theta}^{(p)}_{(C)}(\mathbf{y}_{s \, gen}^{(g,b)}) 
\end{matrix} \right]^T,$ for $p=1,2,\dots,P$, where $g=1,2,\dots, G$ and $b=1,2,\dots, B$. The output of the second step of the procedure are vectors of simulated ex ante prediction errors $\hat{\pmb\theta}^{(p)}(\mathbf{y}_{s \, gen}^{(g,b)}) - \pmb\theta(\mathbf{y}_{gen}^{(g,b)})$, each of size $C \times 1$, for $P$ prediction strategies, where the number of these vectors equals $B \times G \times P$.

\subsection{Ex ante prediction accuracy assessment and voting algorithms} \label{subsec. voting}
Let us now address the fourth step in the WASP method, which comprises points 6 and 7 presented in Figure \ref{fig_scheme}. In point 6 $M$ accuracy measures are computed based on simulated ex ante prediction errors for each prediction strategy. Accuracy measures can be defined in many ways, for example, the prediction RMSE, given by (\ref{RMSE}), and QAPE of different orders, given by (\ref{QAPE}). We write their values as a matrix with $S = G \times C \times M$ rows and $P$ columns, where $G$ is the number of data generation models, $C$ is the number of predicted characteristics, $M$ is the number of considered accuracy measures and $P$ is the number of considered prediction strategies. We refer to this matrix as the accuracy measures matrix and denote by $\mathbf{A}$. 

In point 7, we choose one out of $P$ prediction strategies applying the most popular voting systems. 
The comprehensive overview of voting systems is considered e.g. in \cite{brandt_handbook_2016}. In the WASP method, we choose four systems in which the input information is the matrix $\mathbf{A}$. They differ in terms of:
\begin{itemize}
    \item various proposals of transformations of the obtained accuracy measures matrix $\mathbf{A}$
    \item different criteria of selection.
\end{itemize}
New transformed matrix is called the voting matrix and here is denoted by $\mathbf{W}$ (due to the mentioned similarity between Table \ref{tab_accuracy measures matrix} and Table \ref{tab_voting matrix}).  

Our first proposal is to mimic the first-past-the-post voting algorithm, see, for example, \citet{felsenthal_review_2012}, where voters choose a single candidate, and the candidate with the highest number of votes wins the election. This system is used, for example, to elect the Parliament of the United Kingdom and its predecessors since the Middle Ages, as well as in parliamentary elections in many other countries. Its application for our problem is presented as Algorithm \ref{FPPV}.

\begin{algorithm}
\caption{The proposed first-past-the-post voting algorithm of the prediction strategy selection}\label{FPPV}
\begin{algorithmic}[1]
\Statex 
The input is an accuracy measure matrix $\mathbf{A}$.  
\For{$s$ in 1:$S$} 
\State \multiline{%
In the $s$th row of $\mathbf{A}$, assign rank 1 for the prediction strategy with the minimum value of $m$th accuracy measure and rank 0 for the rest of the strategies.}
\EndFor
\State Write the resulting ranks as a voting matrix $\mathbf{W_1}$ with $S$ rows and $P$ columns.
\State Compute column sums of the ranks for each out of $P$ potential prediction strategies.
\State The prediction strategy with the highest sum is chosen. 
\end{algorithmic}
\end{algorithm}

Our second proposal mimics the Borda count algorithm, which belongs to positional voting procedures. It is discussed among others in \citet[][p.~26]{felsenthal_electoral_2012}. The idea is that voters order candidates from the worst to the best, and in the original Borda count algorithm, the candidate with the most points is elected. Many positional voting algorithms exist where the differences are due to the criterion of election. Currently, one of them called the alternative vote \citep[see][p.~26 for details]{felsenthal_electoral_2012}, is used in elections of the president of the Republic of Ireland, elections to the Australian House of Representatives, and some municipal elections in the United States. However, in our proposal presented as Algorithm \ref{Borda_algo}, we do not use the sum (or equivalently, the mean) of the ranks as in the Borda count algorithm, but the median because ranks are measured on the original scale.

\begin{algorithm}
\caption{The proposed positional voting algorithm of the prediction strategy selection}\label{Borda_algo}
\begin{algorithmic}[1]
\Statex The input is an accuracy measure matrix $\mathbf{A}$.  
\For{$s$ in 1:$S$} 
\State \multiline{%
In the $s$th row of $\mathbf{A}$, rank prediction strategies according to the values of the $m$th accuracy measure from 1 (the maximum value in this row) to $P$ (the minimum value in this row).}
\EndFor
\State Write resulting ranks as a voting matrix $\mathbf{W_2}$ with $S$ rows and $P$ columns.
\State Compute the median of ranks for each out of $P$ prediction strategies.
\State The prediction strategy with the highest median rank is chosen. 
\end{algorithmic}
\end{algorithm}

The proposed third method is inspired by two algorithms: evaluative voting \citep[also called utilitarian voting, see][]{baujard_whos_2014} and Majority Judgement \citep{balinski_theory_2007}. In evaluative voting, each voter does not rank candidates but instead evaluates all candidates by assigning an ordinal grade that reflects their suitability. Usually, it is assumed that the better the candidate the higher the grade. Hence, unlike in the score voting, the same grade can be given to several candidates. The candidate with the highest total (or equivalently -- the average) grade is declared the winner. The procedure of the Majority Judgement is very similar except for the fact the highest median grade (instead of the sum or mean) is the rule to choose the candidate. Our method shares the improvement of the evaluative voting and Majority Judgement compared with the score algorithm, that judging on grades, there is no dependency on alternatives \citep[compare][p.~8]{kemm_why_2023}. The key distinction between these two algorithms and our proposed procedure is that the assessment in our approach is based on scaled accuracy measures (and, therefore, results measured on a ratio scale) rather than the ranks (and therefore, results measured on the ordinal scale). This is a significant advantage of the proposed approach, presented as Algorithm \ref{voting_algo}. To scale the values of accuracy measures, we propose to use the following formula:
\begin{equation}\label{scaling}
a'= 1 - a,    
\end{equation}
where $a$ is a rescaled value (min-max normalisation) of the considered accuracy measure. It also leads to values from the interval $[0,1]$. However, in our case this transformation has the following advantage resulting in its interpretation. Similarly to ranks in Algorithm \ref{Borda_algo} and the utility function, the higher the value of (\ref{scaling}), the better is accuracy.

\begin{algorithm}\caption{The proposed evaluative voting algorithm of the prediction strategy selection}\label{voting_algo}
\begin{algorithmic}[1]
\Statex The input is an accuracy measure matrix $\mathbf{A}$.  
\For{$s$ in 1:$S$} 
\State \multiline{%
In the $s$th row of $\mathbf{A}$, scale the values of the $m$th accuracy measure applying formula (\ref{scaling}) to obtain $P$-values from interval $[0,1]$.}
\EndFor
\State Write resulting values as a matrix $\mathbf{W_3}$ with $S$ rows and $P$ columns (called the voting matrix).
\State
Compute the median of scaled values for each out of $P$ prediction strategies.
\State the prediction strategy with the highest value of the median is chosen. 
\end{algorithmic}
\end{algorithm}

The drawback of Algorithm \ref{voting_algo} is that it only considers the median of the results in the selection process. Although the median can be replaced by a different quantile, the comparison takes into account only a single point of the distribution. As a solution, we propose the following Algorithm \ref{ECDF_algo}, in which the choice of the prediction strategy is based on the voting matrix $\mathbf{W_3}$ defined in Algorithm \ref{voting_algo}, but with a different selection criterion. 
We interpret these values as realisations of the random variable corresponding to the scaled accuracy measures. It allows us to further estimate the empirical c.d.f (ECDF). It is important to note that the values in each column of this matrix fall within the interval $[0,1]$. Therefore, the area under a curve representing the ECDF (AUC ECDF) of the values in the $p$th column of $\mathbf{W_3}$ matrix within the interval $[0,1]$ also lies within the interval $[0,1]$. The area equal to $0$ is achieved for the $p$th prediction strategy when all values in the $p$th column of the $\mathbf{W_3}$ matrix equal $1$. It can be observed in the case of the prediction strategy, which performs the best in all the cases considered in the simulation study. Therefore, in Algorithm \ref{ECDF_algo}, we propose to choose the prediction strategy with the smallest value of the AUC ECDF of the scaled accuracy measures in the interval $[0,1]$. 

Finally, the similarity of the Algorithm \ref{ECDF_algo} to the stochastic dominance rules \citep[definitions are presented for example in][p.~252]{levy_stochastic_1990} can be observed. It leads to the following implication. If the values of $\mathbf{W_3}$ for the $p$th predictor have first-order (or second-order) stochastic dominance over the values of $\mathbf{W_3}$ for the $p'$th predictor, then the predictor is better in the sense of the Algorithm \ref{ECDF_algo}.  

\begin{algorithm}
\caption{The proposed ECDF AUC-based voting algorithm of the prediction strategy selection}\label{ECDF_algo}
\begin{algorithmic}[1]
\Statex The input is an accuracy measure matrix $\mathbf{A}$.  \For{$s$ in 1:$S$} 
\State \multiline{%
In the $s$th row of $\mathbf{A}$, based on formula (\ref{scaling}) scale the values of the $m$th accuracy measure in this row to obtain $P$-values from interval $[0,1]$.}
\EndFor
\State Write resulting values as a matrix $\mathbf{W_3}$ with $S$ rows and $P$ columns (called the voting matrix).
\State
Based on the values in the $p$th column, compute the empirical cumulative distribution function (ECDF) for the scaled voting results obtained for the $p$th prediction strategy (where $p = 1, 2, \dots, P$).
\State The prediction strategy with the smallest value of the area under the ECDF in interval $[0,1]$ is chosen. 
\end{algorithmic}
\end{algorithm}
It should be noted that the maximum area is $1$. What is more, for the hypothetical prediction strategy which is the best for all considered cases (with 1s in a specific column of the voting matrix), the area under its ECDF in this interval is $0$. For strategies which are not the best in all cases, the area is higher than $0$ (and smaller or equal 1).

\section{Automobile portfolio example}
One area of extensive use of predictive models is in the automobile insurance industry and more specifically, in the problem of insurance portfolio management. An insurance portfolio can be defined as a collection of insurance policies. A set premium is charged for each policy within the portfolio. However, future claims of an unknown amount contribute to the portfolio's uncertainty. To manage this uncertainty, risk measures such as value-at-risk (VaR) or tail value-at-risk (TVar) are traditionally used \cite{denuit2006actuarial}. These measures typically summarize a random variable's distribution function as a single number. In the context of an insurance portfolio, this random variable is defined as the aggregate value of claims for a single policy. Another alternative, particularly in the era of big data, is to use statistical modeling. It allows estimating, for instance, the expectation of the aggregate value of claims for a single policy or the quantile, while taking into account the risk factors that affect claims. 
As the portfolio observes mostly renewable policies over successive periods, it seems to be reasonable not only to estimate but also to predict measures to support the risk management process. When evaluating the entire portfolio, the future aggregate value of claims for the portfolio, as well as other characteristics such as the median or the mean can be predicted. In this scenario, the primary concern of selecting an appropriate prediction strategy arises. Undertaking a prediction for a future period suggests using an ex ante approach, which is inherently future-oriented. That is why we introduce the implementation of the WASP method, as in Figure \ref{fig_scheme}, in the insurance portfolio management. The WASP method enables the simultaneous selection of the winning prediction strategy for both the total and median of the aggregate value of claims, according to the value of ex ante prediction accuracy measures. It also permits the integration of various types of accuracy measures, e.g. $RMSE$, $MAE$, $QAPE$, which serves to enhance the precision of our predictions. Furthermore, since the prediction is for a future period, an ex ante approach is adopted.

We carry out an empirical study based on the real automobile portfolio taken from a Polish insurance company for full 2007-2010 years (exposure = 1). The goal is to identify a prediction strategy that is appropriate for the entire portfolio in 2011. The input dataset consists 20,734 policies and among the policies we count 1,565 claims. Generally, the most common risk factors describe: the policyholder, the subject of insurance and a geographical variable. In our data, each $i$th policy corresponds to the aggregate value of claims for a single policy $Claim\_Amount$ and in addition, each policy is assigned risk factor categories:
\begin{itemize}
\item \textit{Gender} -- category: 1 (Female), 0 (Male)
\item \textit{Kind\_of\_distr} -- kind of district: urban, country, suburban
\item \textit{Kind\_of\_payment} -- category: cash, transfer
\item \textit{Engine} -- category: BEN, DIE
\item \textit{Age\_group} -- category: 1, 2, 3.
\end{itemize}
In the predictive model, the variable $Claim\_Amount$ is assumed as the dependent variable while risk factors as independent variables. Two key characteristics describing the policy portfolio are taken into account in the prediction process: the total value of claims denoted as $\theta_{(1)}$ and the median of claims denoted as $\theta_{(2)}$. Table \ref{tab_portfolio} presents the values of the characteristics over the years. 
\begin{table}
\caption{Values of predicted characteristics in past periods in PLN}
\centering
\begin{tabular}{rrrrr} 
  \hline
 & Year & n & Total\_claim & Median\_claim \\
  \hline
1 & 2007 & 337 & 1,232,108.93 & 2,566.50 \\
  2 & 2008 & 356 & 1,268,510.54 & 2,294.51 \\
  3 & 2009 & 385 & 1,159,533.04 & 2,060.79 \\
  4 & 2010 & 487 & 1,474,269.36 & 2,357.60 \\
   \hline
\end{tabular}\label{tab_portfolio}
\end{table}

\begin{figure}[H]\label{fig_dist1}
  \centering
  \includegraphics[scale = 0.65]{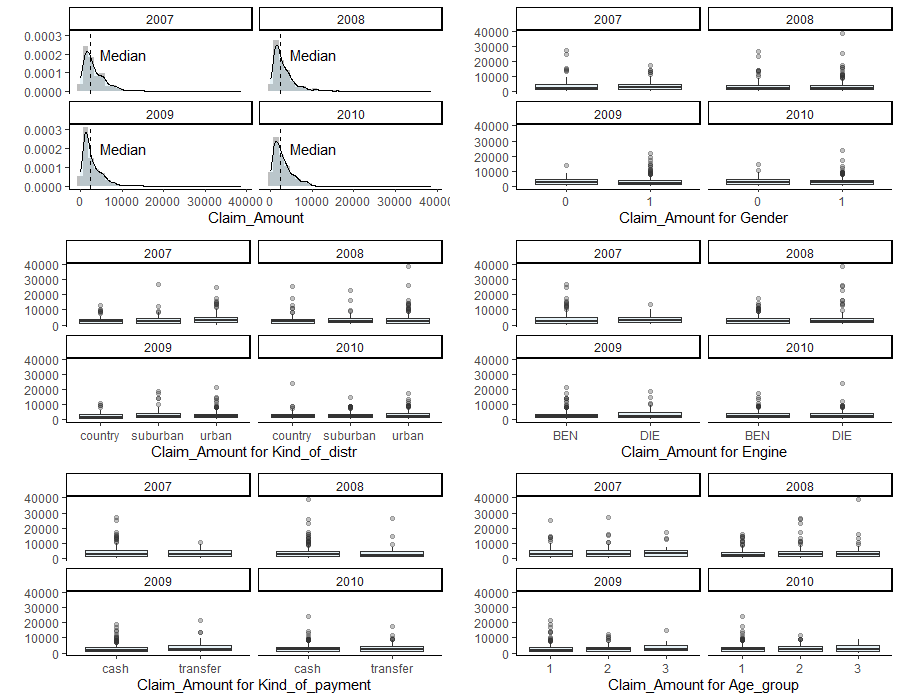}
  \caption{The histogram of the non-zero aggregate value of claims for single policy (left top panel), the box plots of the non-zero aggregate value of claims for single policy in risk groups (other panels)}\label{dist1}
\end{figure}

There is a moderate variation between the values in the following years. In turn, Figure \ref{fig_dist1} illustrates a data analysis exploring risk factors that influence claim amounts in the entire portfolio.

It depicts a series of plots that visualize the distribution of non-zero aggregate value of claims across years and different categorical variables in years 2007, 2008, 2009, and 2010. The top left histogram shows the distribution of the value of claims. The histogram is highly skewed right, with most claims being relatively small, but some very large claims are present. Other plots show box plots of the value of claims, separated by different risk factors. The median claim amount for women is slightly higher than for men, but the overall distributions are very similar. Urban areas tend to have higher claims compared to rural areas, and the median for DIE engines appears to be slightly higher than that of BEN engines. There is also a greater number of outlier claims for DIE engines. The distribution of claim values appears to be similar for both cash and transfer payments, and there is an insignificant trend for claims to increase with age group.

In order to identify a strategy that is appropriate for predicting the total and median of the aggregate value of claims for the entire portfolio in 2011 we apply the WASP method. It requires a number of important assumptions to be made. The first assumption relates to how the value of the variable $ Claim\_Amount$ is generated to obtain ‘true’ claims values. Generation from a statistical model is assumed and an ensemble of 6 models most popular in insurance portfolio management is used:
\begin{itemize}
\item parametric $M_{PAR}$: GLM Gamma (GG), log-normal regression (LogN), GAM (see \cite{de2008generalized}, \cite{denuit2004non}, \cite{frees2009regression}),
\item non-parametric $M_{NPAR}$: decision tree (DT), SVM with linear kernel (SVML), SVM with polynomial kernel (SVMP) (see \cite{wuthrich2020bias}, \cite{yu2021claim}, \cite{matthews2022machine}).
\end{itemize}
In the second assumption, the predictive models used to predict characteristics $\theta_{(1)}$ and $\theta_{(2)}$ should be adopted. We assume the same ensemble GG, LogN, GAM, DT, SVML, and SVMP  as in the generation process. The third assumption points a prediction algorithm for every predictive model. We apply the most popular $PLUG-IN$ algorithm for every model. 
Since we assume six predictive models and one prediction algorithm throughout the WASP method, we choose one of six strategies presented in Table \ref{strat}. 
\begin{table}[H]
\caption{Description of candidate prediction strategies}
\begin{center}
\begin{tabular}{c | c   c c | } \label{strat}
 Strategy & Predictive model & Prediction algorithm \\ \hline
 strategy1 & GG & PLUG-IN \\
 strategy2 & LogN & PLUG-IN \\
 strategy3 & GAM & PLUG-IN \\
 strategy4 & DT & PLUG-IN \\
 strategy5 & SVML & PLUG-IN \\
 strategy6 & SVMP & PLUG-IN \\
\hline
\end{tabular}
\end{center}
\end{table}
The fourth assumption relates to the criteria for selecting a prediction strategy. We employ three accuracy measures: $RMSE$, $QAPE_{0.5}$ and $QAPE_{0.95}$.  

After establishing the input assumptions of the WASP method, we conduct a Monte Carlo simulation with 5000 iterations. The single $b$th iteration has the following stages:
\begin{enumerate}
\item generating real values of the variable of $Claim\_Amount$ under 6 assumed models $M_{(1)}, ..., M_{(6)}$,
\item calculating characteristics $\theta_{(1)}$ and $\theta_{(2)}$ for every generated ‘true’ $Claim\_Amount$,
\item predicting characteristics $\theta_{(1)}$ and $\theta_{(2)}$ under 6 assumed models $M_{(1)}, ..., M_{(6)}$,
\item calculating measures of ex ante prediction accuracy: $RMSE$, $QAPE_{0.5}$, $QAPE_{0.95}$.
\end{enumerate}
We therefore obtain, on the one hand, the future ‘true’ values of the characteristics that the Monte Carlo experiment generates and, on the other hand, their predictions. This allows us to assess the ex ante prediction accuracy and create the accuracy measures matrix $\mathbf{A}$. In our example, this matrix has $36$ rows, as we involve two characteristics, six models and three accuracy measures, see Table \ref{tab_sim_results}. 

\begin{table}
\caption{Accuracy measures matrix $\mathbf{A}$ -- the short form}
\begin{center}
\begin{tabular}{c | c c c c c c | } \label{tab_sim_results}
 voter  & strategy1 & strategy2 & strategy3 & strategy4 & strategy5 & strategy6 \\
\hline
{$M_1, \space \theta_1, \space RMSE$} & 333,175 & 634,604 & 333,481 & 502,998 & 590,183 & 512,242 \\
\dots  & \dots & \dots & \dots &  \dots & \dots &  \dots  \\
{$M_6, \space \theta_1, \space RMSE$} & 460,624 & 707,067 & 460,902 & 691,018 & 688,544 & 707,864\\
{$M_1, \space \theta_2, \space RMSE$} & 935 & 793 & 932 & 708 & 823 & 696  \\
\dots  & \dots & \dots & \dots &  \dots & \dots &  \dots  \\
{$M_6, \space \theta_2, \space RMSE$} & 1,229 & 750 & 1,229 & 753 & 832 & 742  \\
{$M_1, \space \theta_1, \space QAPE_{0.5}$} & 218,998 & 509,506 & 220,136 & 342,380 & 439,140 & 358,066 \\
\dots  & \dots & \dots & \dots &  \dots & \dots &  \dots  \\
{$M_6, \space \theta_1, \space QAPE_{0.5}$} & 284,903 & 470,012 & 283,133 & 447,735 & 436,563 & 473,420 \\
{$M_1, \space \theta_2, \space QAPE_{0.5}$} & 755 & 472 & 752 & 447 & 517 & 455 \\
\dots  & \dots & \dots & \dots &  \dots & \dots &  \dots  \\
{$M_6, \space \theta_2, \space QAPE_{0.5}$} & 1,086 & 467 & 1,084 & 478 & 519 & 465\\
{$M_1, \space \theta_1, \space QAPE_{0.95}$} & 638,488 & 1,130,219 & 642,936 & 973,300 & 1,104,696 & 974,396 \\
\dots  & \dots & \dots & \dots &  \dots & \dots &  \dots  \\
{$M_6, \space \theta_1, \space QAPE_{0.95}$} & 854,513 & 1,363,886 & 854,983 & 1,341,894 & 1,337,557 & 1,357,542\\
{$M_1, \space \theta_2 \space QAPE_{0.95}$} & 1,669 & 1,598 & 1,668 & 1,353 & 1,641 & 1,327\\
\dots  & \dots & \dots & \dots &  \dots & \dots &  \dots \\
{$M_6, \space \theta_2, \space QAPE_{0.95}$} & 2,051.54 & 1,483 & 2,050 & 1,480 & 1,670 & 1,484 \\
 \hline
\end{tabular}
\end{center}
\end{table}

Then, we use this matrix to conduct a voting, in which $36$ voters determine the winner of six candidate strategies. The WASP method proposes four voting systems which can be used. In each system, the initiating step is to perform some transformation of the matrix $\mathbf{A}$ to the voting matrix. 

The first voting corresponds to FPTP system (see Algorithm \ref{FPPV}). The smallest value of accuracy measure in each row of the matrix $\mathbf{A}$ is replaced by $1$ and all other values by $0$. It leads to the voting matrix $\mathbf{W_1}$ presented in Table \ref{tab_W1}.

\begin{table}
\caption{ First-past-the-post voting matrix $\mathbf{W_1}$ -- the short form} 
\begin{center}
\begin{tabular}{c | c c c c c c | } \label{tab_W1}
 voter & strategy1 & strategy2 & strategy3 & strategy4 & strategy5 & strategy6 \\
\hline
{$M_1, \space \theta_1, \space RMSE$} &1  & 0  & 0  & 0  & 0  & 0  \\
\dots  & \dots & \dots & \dots &  \dots & \dots &  \dots  \\
{$M_6, \space \theta_1, \space RMSE$} & 1  & 0  & 0  & 0  & 0  & 0 \\

{$M_1, \space \theta_2, \space RMSE$} &  0  & 0  & 0  & 0  & 0  & 1  \\
\dots  & \dots & \dots & \dots &  \dots & \dots &  \dots  \\
{$M_6, \space \theta_2, \space RMSE$} & 0  & 0  & 0  & 0  & 0  & 1    \\

{$M_1, \space \theta_1, \space QAPE_{0.5}$} &1  & 0  & 0  & 0  & 0  & 0  \\
\dots  & \dots & \dots & \dots &  \dots & \dots &  \dots  \\
{$M_6, \space \theta_1, \space QAPE_{0.5}$} & 0  & 0  & 1  & 0  & 0  & 0  \\

{$M_1, \space \theta_2, \space QAPE_{0.5}$} & 0  & 0  & 0  & 1  & 0  & 0  \\
\dots  & \dots & \dots & \dots &  \dots & \dots &  \dots  \\
{$M_6, \space \theta_2, \space QAPE_{0.5}$} & 0  & 0  & 0  & 0  & 0  & 1 \\

{$M_1, \space \theta_1, \space QAPE_{0.95}$} & 1  & 0  & 0  & 0  & 0  & 0 \\
\dots  & \dots & \dots & \dots &  \dots & \dots &  \dots  \\
{$M_6, \space \theta_1, \space QAPE_{0.95}$} & 1  & 0  & 0  & 0  & 0  & 0 \\

{$M_1, \space \theta_2 \space QAPE_{0.95}$} & 0  & 0  & 0  & 0  & 0  & 1  \\
\dots  & \dots & \dots & \dots &  \dots & \dots &  \dots \\
{$M_6, \space \theta_2, \space QAPE_{0.95}$} & 0 & 0  & 0  & 1  & 0  & 0 \\
 \hline
sum & 12  & 4  & 11  & 3  & 0  & 6 \\
\end{tabular}
\end{center}
\end{table}

For example, the smallest value in the first row of $\mathbf{W_1}$ corresponds to strategy 1, and hence, it shows a value of 1 in the first column. Then, the sums of votes for each candidate prediction strategy are computed. The highest value of the sum is 12, which indicates strategy 1 as the winning strategy. The second voting is conducted in accordance with the positional voting system (see Algorithm \ref{Borda_algo}). In the matrix $\mathbf{A}$, each row is replaced by ranks from $1$ to $6$, resulting in the voting matrix $\mathbf{W_2}$ as outlined in Table \ref{tab_W2}. 

The higher the rank, the better the prediction accuracy, with the smallest accuracy measure values indicating the highest rank. For example, the smallest value in the first row of the matrix $\mathbf{A}$ is observed for strategy 1, and hence, in the first row of $\mathbf{W_2}$, the highest rank equal to $6$ is in the first column. Thereafter, the medians of ranks are computed for every candidate strategy. The highest value of the median of ranks is equal to $5$. This points to strategy 1 and at the same time strategy 3 as the winning strategy. Due to the fact that ultimately each strategy is assigned a rank as an integer, it may happen that several strategies obtain the maximum value of the selection criterion. There is then no clear winner. In the third and fourth voting, we involve the evaluative system (see Algorithm \ref{voting_algo}) and the ECDF AUC system (see Algorithm \ref{ECDF_algo}). The same voting matrix $\mathbf{W_3}$ presented in Table \ref{tab_W3} is provided as input. 

\begin{table}
\caption{ Positional voting matrix $\mathbf{W_2}$ -- the short form}
\begin{center}
\begin{tabular}{c | c c c c c c | } \label{tab_W2}
 voter & strategy1 & strategy2 & strategy3 & strategy4 & strategy5 & strategy6 \\
\hline
{$M_1, \space \theta_1, \space RMSE$} & 6 &   1 &   5 &   4 &   3 &   2 \\
\dots  & \dots & \dots & \dots &  \dots & \dots &  \dots  \\
{$M_6, \space \theta_1, \space RMSE$} &  6 &   2 &   5 &   3 &   4 &   1\\

{$M_1, \space \theta_2, \space RMSE$} &  1 &   4 &   2 &   5 &   3 &  6\\
\dots  & \dots & \dots & \dots &  \dots & \dots &  \dots  \\
{$M_6, \space \theta_2, \space RMSE$} & 2 &   5 &   1 &   4 &   3 &   6 \\

{$M_1, \space \theta_1, \space QAPE_{0.5}$} &  6 &   1 &   5 &   4 &   2 &   3\\
\dots  & \dots & \dots & \dots &  \dots & \dots &  \dots  \\
{$M_6, \space \theta_1, \space QAPE_{0.5}$} & 5 &  2 &  6 &   3 &   4 &  1  \\

{$M_1, \space \theta_2, \space QAPE_{0.5}$} & 1 &   4 &   2 &   6 &   3 &   5 \\
\dots  & \dots & \dots & \dots &  \dots & \dots &  \dots  \\
{$M_6, \space \theta_2, \space QAPE_{0.5}$} & 1 &   5 &   2 &   4 &   3 &   6\\

{$M_1, \space \theta_1, \space QAPE_{0.95}$} & 6 &   1 &   5 &   4 &  2 &  3 \\
\dots  & \dots & \dots & \dots &  \dots & \dots &  \dots  \\
{$M_6, \space \theta_1, \space QAPE_{0.95}$} &  6 &   1 &   5 &   3 &   4 &  2\\

{$M_1, \space \theta_2 \space QAPE_{0.95}$} & 1 &   4 &   2 &  5 &   3 &   6  \\
\dots  & \dots & \dots & \dots &  \dots & \dots &  \dots \\
{$M_6, \space \theta_2, \space QAPE_{0.95}$} & 1 &   5 &  2 &  6 &   3 &  4 \\
 \hline
median & 5  & 4  & 5  & 3  & 3  & 3 \\
\end{tabular}
\end{center}
\end{table}

\begin{figure} \label{Fig_ECDF}
	\begin{center}
		\includegraphics[width=16cm]{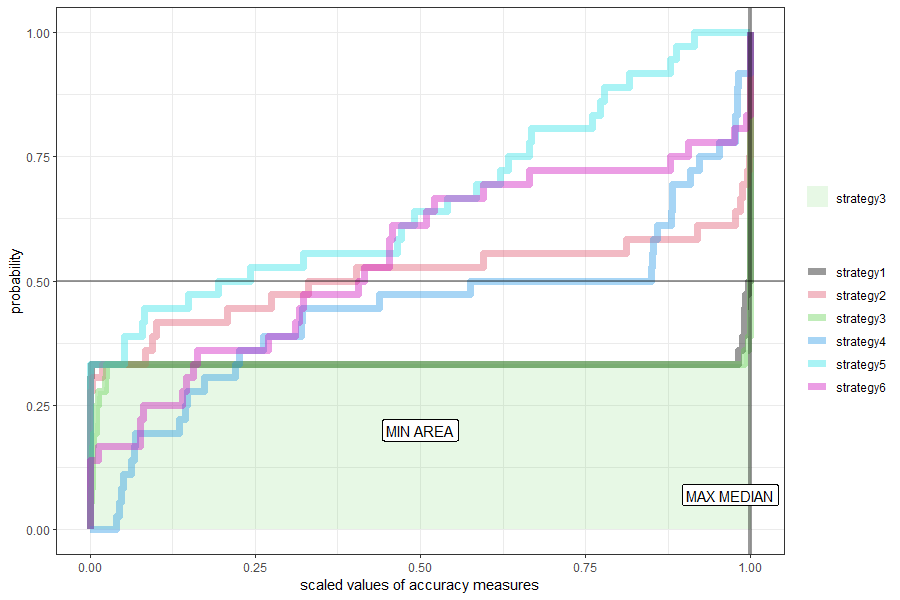}
		\caption{Results of ECDF AUC voting system}
  	\end{center}
\end{figure}

This matrix is computed by scaling values by rows of accuracy measures matrix $\mathbf{A}$ according to the formula (\ref{scaling}). Considering, for example, the first row of accuracy matrix $\mathbf{A}$, the smallest value (the best accuracy) is observed for strategy 1, and hence, after scaling, in the first row of matrix  $\mathbf{W_3}$ (see Table \ref{tab_W3}), 1 is obtained in the first column. It can be seen that there happen to be more zeros and ones in single rows. The reason for this is that similar accuracy measures come out in identical scaled values when rounding to too few decimal places. This is easily noticed by comparing matrices $\mathbf{A}$ and $\mathbf{W_3}$ (see Tables \ref{tab_sim_results} and \ref{tab_W3}). Thereafter, in evaluative algorithm \ref{voting_algo}, the median of scaled values for each strategy is determined and the strategy with the highest median value wins. In turn, ECDF AUC voting system starts with obtaining the empirical c.d.f of scaled accuracy measure for every candidate strategy. Then, the area under every ECDF is computed. The strategy with minimum ECDF AUC is the winning one. Figure \ref{Fig_ECDF} presents ECDFs of candidate strategies and the smallest ECDF AUC points the winner, which is strategy 3.

\begin{table}
\caption{Scaled voting matrix $\mathbf{W_3}$ -- the short form}
\begin{center}
\begin{tabular}{c | c c c c c c | } \label{tab_W3}
 voter & strategy1 & strategy2 & strategy3 & strategy4 & strategy5 & strategy6 \\
\hline
{$M_1, \space \theta_1, \space RMSE$} & 1.000 & 0.000 & 0.999 & 0.437 & 0.147 & 0.406   \\
\dots  & \dots & \dots & \dots &  \dots & \dots &  \dots  \\
{$M_6, \space \theta_1, \space RMSE$} &  1.000 & 0.003 & 0.999 & 0.068 & 0.078 & 0.000 \\
{$M_1, \space \theta_2, \space RMSE$} & 0.000 & 0.594 & 0.009 & 0.953 & 0.470 & 1.000  \\
\dots  & \dots & \dots & \dots &  \dots & \dots &  \dots  \\
{$M_6, \space \theta_2, \space RMSE$} & 0.001 & 0.984 & 0.000 & 0.978 & 0.817 & 1.000   \\
{$M_1, \space \theta_1, \space QAPE_{0.5}$} & 1.000 & 0.000 & 0.996 & 0.575 & 0.242 & 0.521  \\
\dots  & \dots & \dots & \dots &  \dots & \dots &  \dots  \\
{$M_6, \space \theta_1, \space QAPE_{0.5}$} & 0.991 & 0.018 & 1.000 & 0.135 & 0.194 & 0.000   \\
{$M_1, \space \theta_2, \space QAPE_{0.5}$} & 0.000 & 0.919 & 0.012 & 1.000 & 0.772 & 0.974  \\
\dots  & \dots & \dots & \dots &  \dots & \dots &  \dots  \\
{$M_6, \space \theta_2, \space QAPE_{0.5}$} & 0.000 & 0.997 & 0.005 & 0.979 & 0.913 & 1.000  \\
{$M_1, \space \theta_1, \space QAPE_{0.95}$} & 1.000 & 0.000 & 0.991 & 0.319 & 0.052 & 0.317 \\
\dots  & \dots & \dots & \dots &  \dots & \dots &  \dots  \\
{$M_6, \space \theta_1, \space QAPE_{0.95}$} & 1.000 & 0.000 & 0.999 & 0.043 & 0.052 & 0.012  \\
{$M_1, \space \theta_2 \space QAPE_{0.95}$} & 0.000 & 0.207 & 0.002 & 0.922 & 0.081 & 1.000   \\
\dots  & \dots & \dots & \dots &  \dots & \dots &  \dots \\
{$M_6, \space \theta_2, \space QAPE_{0.95}$} & 0.000 & 0.995 & 0.002 & 1.000 & 0.668 & 0.993  \\
 \hline
 median & 0.335  & 0.506  & 0.331  & 0.442  & 0.577  & 0.547 \\
 \end{tabular}
\end{center}
\end{table}

A summary of the results is presented in Table \ref{tab_criteria}. 
\begin{table}[H]
\begin{center}
\begin{threeparttable}
\caption{Values of selection criteria for four voting systems}
\label{tab_criteria}
\begin{tabular}{lrrrr}
\hline
\multicolumn{1}{p{2.5cm}}{\centering Prediction strategy} & \multicolumn{1}{p{2.5cm}}{\centering FPTP \\ voting\tnote{(a)}} \space  & \multicolumn{1}{p{2.5cm}}{\centering Positional \\ voting \tnote{(b)}} \space  & \multicolumn{1}{p{2.5cm}}{\centering Evaluative \\ voting\tnote{(c)}} \space  & \multicolumn{1}{p{2.5cm}}{\centering ECDF AUC  \\ voting \tnote{(d)}}  \\ 
  \hline
 strategy1 &\textbf{12} & \textbf{5} & 0.997 & 0.335 \\
 strategy2 & 4 & 4 & 0.366 & 0.506 \\ 
 strategy3 & 11 & \textbf{5} & \textbf{0.999} & \textbf{0.331} \\ 
 strategy4 & 3 & 3 & 0.712 & 0.442 \\  
 strategy5 & 0 & 3 & 0.217 & 0.577 \\  
 strategy6 & 6 & 3 & 0.410 & 0.547 \\ 
 \hline
\end{tabular}
\begin{tablenotes}
      \small
\item[(a)] sum of votes (the higher the better)
\item[(b)] median of ranks (the higher the better)
\item[(c)] median of scaled accuracy measures (the higher the better)
\item[(d)] ECDF AUC (the smaller the better)
    \end{tablenotes}
    \end{threeparttable}
    \end{center}
\end{table}

Upon analyzing the data in Table \ref{tab_criteria} it can be seen that the indicated strategy is not the same in every voting system. According to the FFTP voting system, strategy 1 should be chosen, based on positional voting  - either strategy 1 or strategy 3. In both evaluative voting and ECDF AUC voting, strategy 3 is the winner, but in these cases, strategy 1 is only slightly worse. However, as discussed in Subsection \ref{subsec. voting}, the preferred voting system is ECDF AUC voting, which takes into account the entire distribution of votes. Therefore, based on ECDF AUC voting, we consider strategy 3 as the winner. Hence, taking GAM as the predictive model together with PLUG-IN algorithm, the prediction for $2011$ amounts to $\theta_{(1)} = 1,445,233.19$ and $\theta_{(2)} = 2,674.49$. 
It is important to note that we are not showing estimates of prediction accuracy measures. While the obvious choice would be to use the values obtained under the predictive model GAM (which is the same as in the winning strategy), the accuracy measures obtained for the winning strategy under different models considered in the WASP method are also entirely valid. The decision, which model should be considered as the source of future variability of the predicted characteristics, depends on the researcher's judgment.
Finally, it is worth noting that we have also studied the distributions of values of the voting matrix using first-order and second-order stochastic dominances. We found that no distribution dominates the others, which means that the application of stochastic dominance for comparison does not provide clear recommendations for selecting a prediction strategy for the considered dataset.

\vskip 3mm

\section{Conclusion} 
The present paper discusses the use of predictive models in economics for decision-making and strategic planning. Specifically, the focus is on regression models for continuous variables and extending the prediction problem to include the joint prediction of any functions of the dependent variable. The distinction between parametric and nonparametric models is highlighted, and the concept of "prediction strategy" is discussed as a pair consisting of the predictive algorithm and the assumed model. 
The primary objective of the proposed simulation-based method is to select the best prediction strategy based on predetermined criteria. 
Based on real data application, the proposed ECDF ROC voting system provides an unambiguous decision on the winning prediction strategy. It considers not only a specific characteristic of the votes (like the sum or median of votes) but their entire distribution. Finally, this approach allows us to predict the considered vector of characteristics and estimate the accuracy based on the same model as used in the winning strategy or any of the models utilized in the WASP algorithm. 

While the proposed method may be time-consuming, it represents an innovative approach with the potential to enhance current business processes. Its properties can enable expansion into new business domains, uncover untapped market opportunities, and contribute to addressing socioeconomic and environmental challenges.

\section*{Declaration of interest}

The authors declare that the research was conducted in the absence of any commercial or financial relationships that could be construed as a potential conflict of interest.

\section*{Research Data Availability Statement}

Due to the sensitive nature of data they cannot be shared.

\section*{Acknowledgement}

The work has been co-financed by the Minister of Science under the "Regional Initiative of Excellence" programme.

\bibliography{WASP}

\newpage

\noindent{Alicja Wolny-Dominiak} \\
Department of Statistical and Mathematical Methods in Economics \\
University of Economics in Katowice, Katowice, Poland\\ 
e-mail: alicja.wolny-dominiak@uekat.pl \\
ORCID: 0000-0002-7484-5538 \\
\url{web.ue.katowice.pl/woali/}

\vspace{0.5cm}

\noindent{Tomasz Żądło} \\
Department of Statistics, Econometrics and Mathematics\\ University of Economics in Katowice, Katowice, Poland \\ 
e-mail: tomasz.zadlo@uekat.pl \\
ORCID: 0000-0003-0638-0748 \\
\url{web.ue.katowice.pl/zadlo/}

\end{document}